\documentclass[3p,times,procedia]{elsarticle}
\usepackage{nupha_ecrc}
\volume{00}
\firstpage{1}
\journalname{Nuclear Physics A}
\runauth{Gojko Vujanovic}
\jid{nupha}
\jnltitlelogo{Nuclear Physics A}
\usepackage{amssymb}
\usepackage[figuresright]{rotating}

\newcommand{\trento}{{\rm T\raisebox{-0.5ex}{R}ENTo}$\,$}

\begin{document}
\begin{frontmatter}
\dochead{XXVIIIth International Conference on Ultrarelativistic Nucleus-Nucleus Collisions\\ (Quark Matter 2019)}
\author{Gojko Vujanovic for the JETSCAPE Collaboration}
\address{Department of Physics and Astronomy, Wayne State University, 666 W. Hancock St, Detroit, Michigan 48201, USA}

\title{Multi-stage evolution of heavy quarks in the quark-gluon plasma}
\begin{abstract}
The interaction of heavy flavor with the quark-gluon plasma (QGP) in relativistic heavy-ion collisions is studied using JETSCAPE, a publicly available software package containing a framework for Monte Carlo event generators. Multi-stage (and multi-model) evolution of heavy quarks within JETSCAPE provides a cohesive description of heavy flavor quenching inside the QGP. As the parton shower develops, a model becomes active as soon as its kinematic region of validity is reached. Two combinations of heavy-flavor energy-loss models are explored within a realistic QGP medium, using parameters which were tuned to describe {\it light-flavor} partonic energy-loss.
\end{abstract}


\end{frontmatter}

\section{Introduction}\label{sec:intro}
JETSCAPE (Jet Energy-loss Tomography with a Statistically and Computationally Advanced Program Envelope) is a modular, flexible, publicly released event-generator framework modeling all aspects of heavy ion collisions. As such, it facilitates a holistic description of heavy-ion collisions, while also allowing the user to focus upon any physics desired by improving/swapping on the corresponding module. There are two main branches of ongoing studies using the JETSCAPE software, both of which are presented in these conference proceedings. The first branch constrains transport properties (such as shear and bulk viscosity) of the quark-gluon plasma (QGP) using a Bayesian comparison of soft (low-$p_T$) hadron data with JETSCAPE simulations \cite{J-F-P_tp}. The other branch focuses on the description of the interaction between hard (high-$p_T$) partons and the QGP. The interaction with the QGP is modeled via a two-step approach describing both light and heavy flavor partons: the high virtuality (and high energy) portion of parton interaction with the QGP is described using  the higher twist formalism \cite{Wang:2001ifa,Majumder:2009ge,Qin:2009gw} implemented in MATTER \cite{Majumder:2013re,Abir:2015hta} (Modular All Twist Transverse-scattering Elastic-drag and Radiation), while the low virtuality (and high energy) quenching of partons is described via Linear Boltzmann Transport (LBT) \cite{Luo:2018pto}. A complete study of high-$p_T$ {\it light flavor} hadronic observables within the JETSCAPE framework is presented in Ref. \cite{A-K_tp}. That study is used in turn to fix all the parameters for heavy-flavors' interaction with the QGP explored within the present contribution, {\it without any additional tuning}. Furthermore, since the heavy quarks' interaction with the QGP using LBT has already been explored in the past \cite{Cao:2018ews}, these proceedings focus more on how MATTER, and the combination of MATTER and LBT, can describe heavy flavor propagation through the QGP.  

\section{Simulation setup}\label{sec:sim}
The event-by-event hydrodynamical simulations used throughout this study are calibrated using an established Bayesian model to data comparison \cite{Bernhard:2019bmu}. That calibration involves simulations using the \trento initial conditions, followed by free-streaming and $(2+1)$-D viscous hydrodynamics. More details can be found in Ref. \cite{Bernhard:2019bmu}. Using the parameters corresponding to the maximum likelihood of the posterior distribution, event-by-event simulations of the QGP medium for Pb-Pb collisions at $\sqrt{s_{NN}}=5.02$ TeV in the 0-10\% centrality class are generated. Hard partons, initially produced via PYTHIA, are deposited into the medium at binary collisions sites given by the \trento initial profile. Those partons are then allowed to exchange their energy and momentum with the pre-simulated dynamically evolving QGP, using either MATTER or LBT as the quenching formalism, depending on the virtuality of each parton.

Based on the higher twist formalism, MATTER is a virtuality-ordered Monte Carlo (MC) event generator describing parton splittings according to a generalized Sudakov form factor, which includes vacuum and in-medium contributions. In this calculation, the in-medium contribution to the Sudakov form factor accounts for transverse momentum broadening of partons ($\hat{q}$) as they travel through the QGP. An effective strong coupling $\alpha_s=0.25$ is used to determine $\hat{q}$, which was tuned using {\it light flavor} observables \cite{A-K_tp}. MATTER proceeds by first sampling the Sudakov form factor in order to determine whether a split has occurred, and if so, calculates the virtuality of the parent parton at the time of its formation. Then, using a medium-modified splitting kernel, the longitudinal momentum fraction of the daughters is determined. In addition to the inelastic splitting process just described, MATTER also includes elastic scatterings with thermal partons in the medium. The scattered thermal partons (also known as recoils) are then transferred back to the JETSCAPE framework, which assigns each parton to the correct module, based on its virtuality (and provided its energy is high enough), for further evolution. Partons with virtuality $Q > Q_s$ (with $Q_s$ being the switching virtuality) are handled by MATTER, while all partons with $Q\leq Q_s$ are given to LBT. The switching virtuality $Q_s$ found to best describe {\it light flavor} observables was $Q_s=2$ GeV. This value of $Q_s$ was also used for heavy flavor evolution within the QGP.             
 
The LBT portion of heavy flavor interaction with the QGP relies on solving the linearized Boltzmann equation, containing $2\to 2$ and $2\to 3$ processes. The $2\to 2$ processes consist of leading order perturbative QCD scatterings between hard and thermal partons. The medium-induced gluon radiation responsible for describing $2\to 3$ processes uses the same higher twist formulation as that employed in MATTER. The resulting recoil particles are treated in the same way as in MATTER. Once partons reach low virtuality and low energy, they are handed back to PYTHIA for hadronization. 

%%%%%%%%%%%%%%%%%%%%%%%%%%%%%%%%%%%%%%%%%%%%%%%%%%%%%%%%%%%%%%%%
\section{Results}\label{sec:results}
%%%%%%%%%%%%%%%%%%%%%%%%%%%%%%%%%%%%%%%%%%%%%%%%%%%%%%%%%%%%%%%%
\begin{figure}[!h]
\begin{center}
\includegraphics[width=0.495\textwidth]{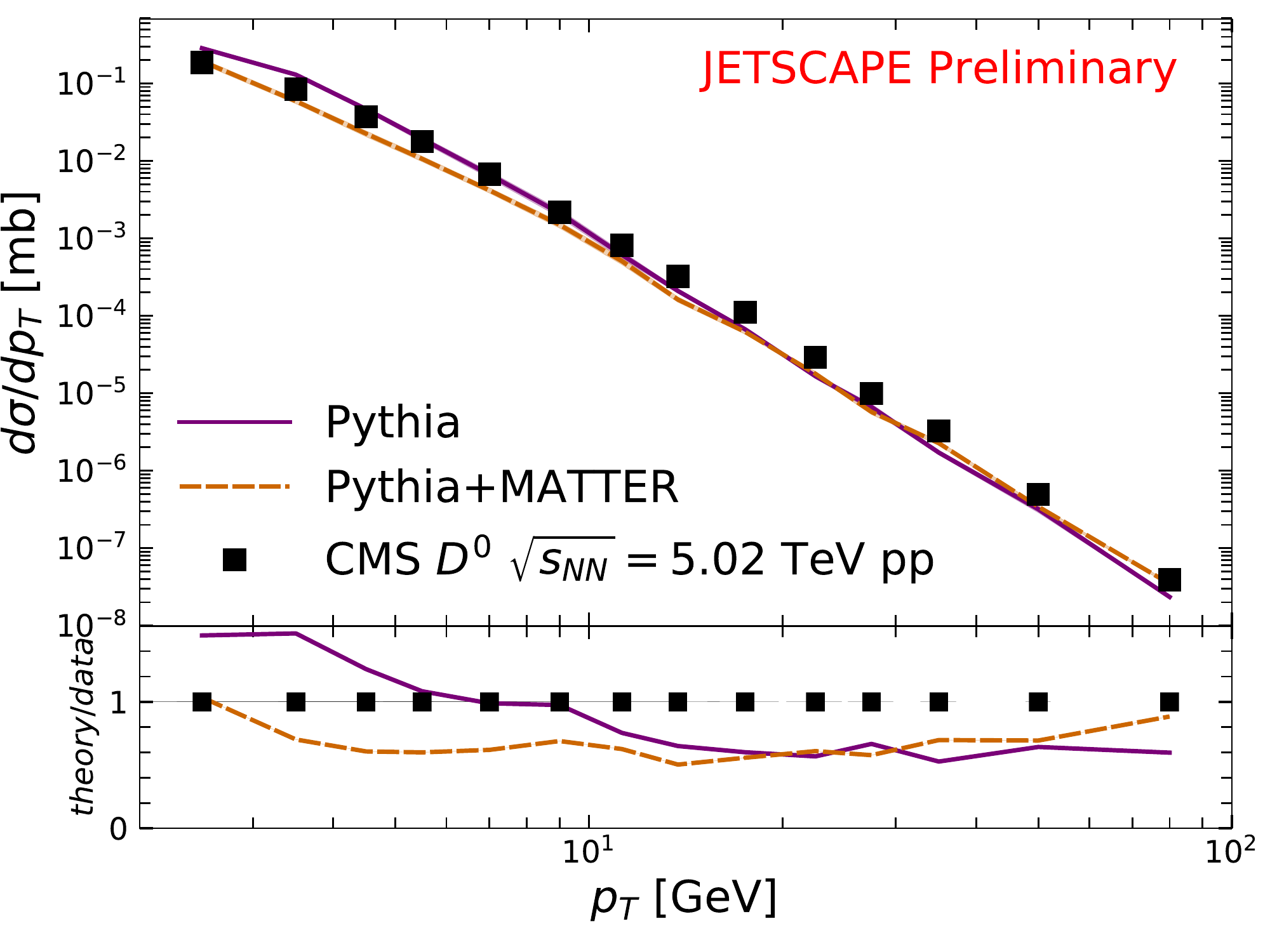}
\includegraphics[width=0.495\textwidth]{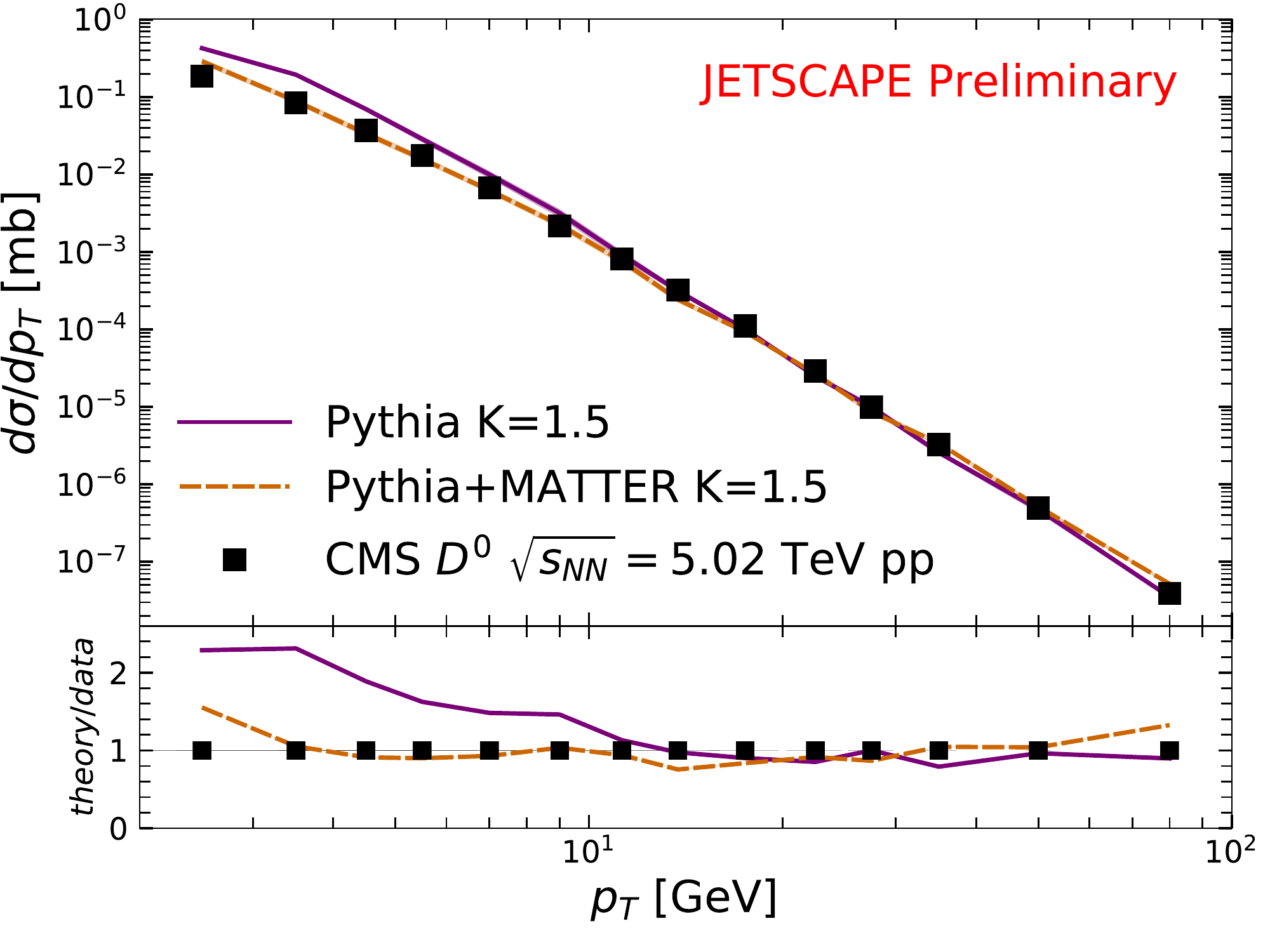}
\end{center}
\caption{(Left) Charm quark production cross section using PYTHIA and a combination of PYTHIA with MATTER. (Right) Same as the left except with a K-factor of 1.5.}
\label{fig:pp}
\end{figure}

The charmed cross section in p+p collisions is explored in Fig. \ref{fig:pp}. The agreement between PYTHIA and CMS data is not as good as when using PYTHIA and MATTER. In proton-proton collisions, it appears that the angular-ordered shower present in PYTHIA generates a charm cross section with a different shape than the CMS data (see left panel of Fig. \ref{fig:pp}). MATTER employs a virtuality ordered shower do describe final state radiation. When used together with PYTHIA,\footnote{When combining PYTHIA and MATTER, PYTHIA is used to produce the hard scattering and initial stage radiation, while MATTER is responsible for the final state showering of the hard partons.} the shape of the charm spectrum is changed and describes the CMS data much better. The theoretical reasons behind this need to be explored further. The right panel of Fig. \ref{fig:pp} highlights better the agreement between our calculations and experimental data, where a K-factor of 1.5 was used to renormalize the charm cross section spectrum. 

\begin{figure}[!h]
\begin{center}
\includegraphics[width=0.495\textwidth]{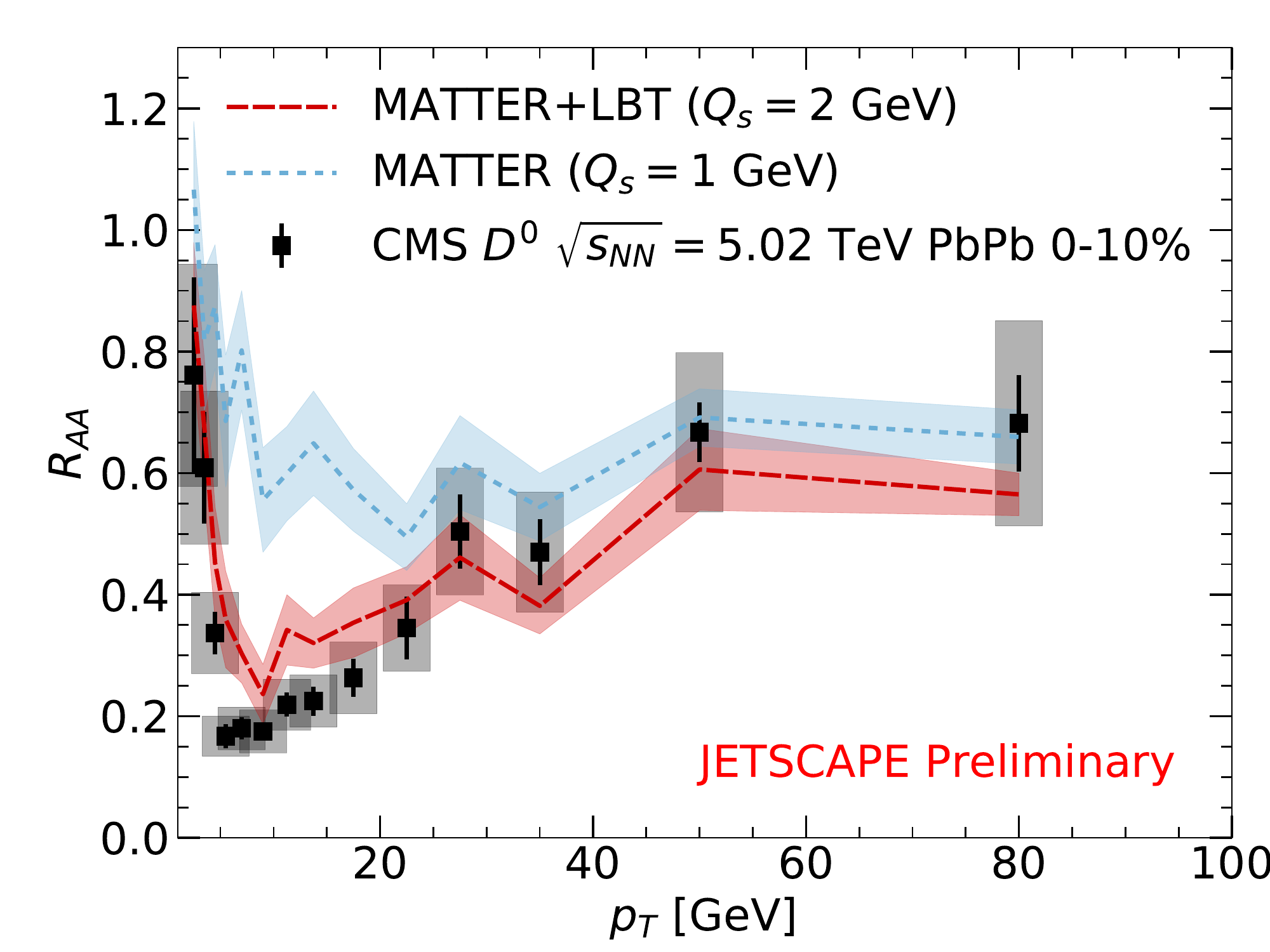}
\includegraphics[width=0.495\textwidth]{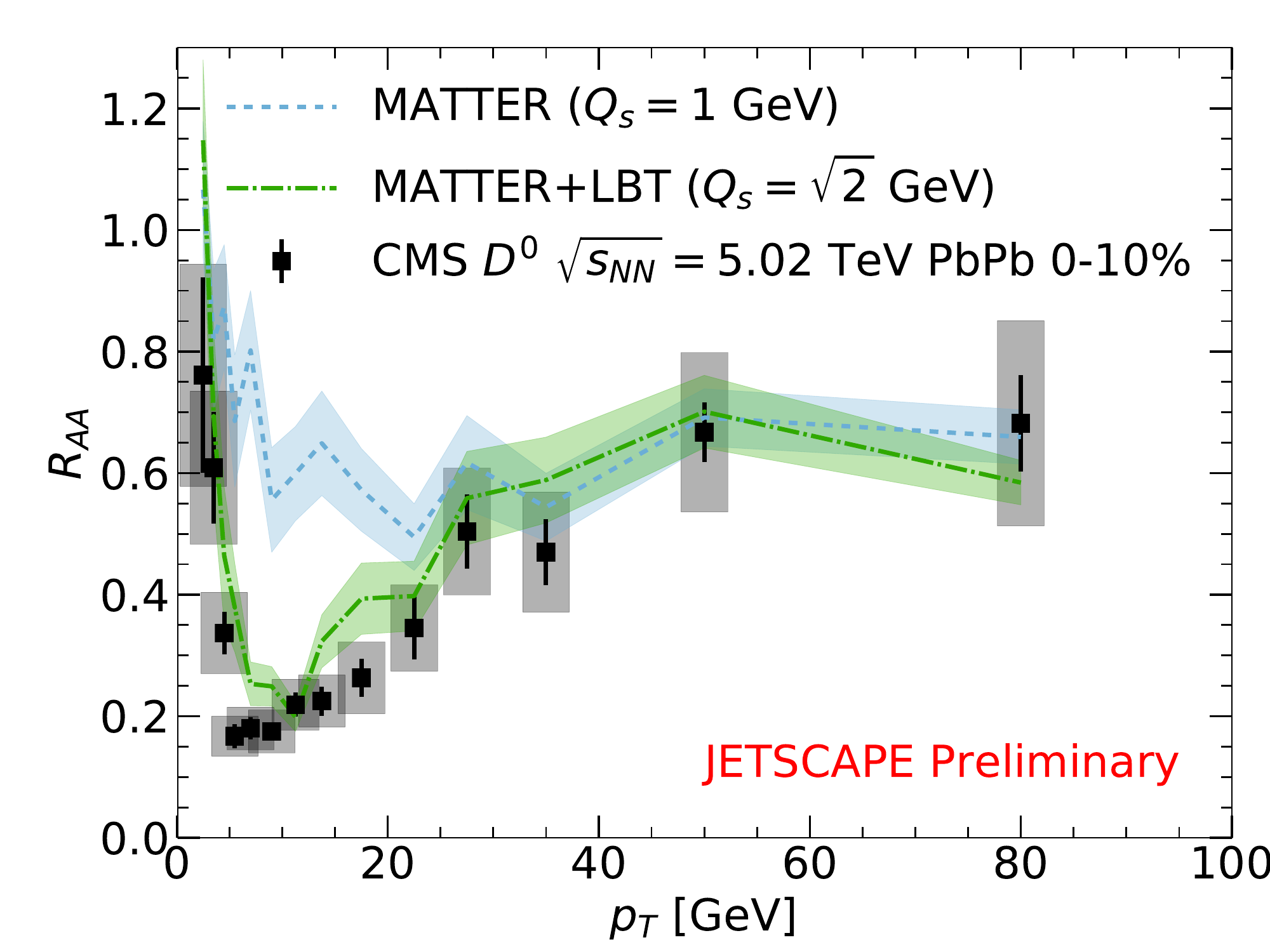}
\end{center}
\caption{(Left) $D^0$ $R_{AA}$ using the JETSCAPE framework and a switching virtuality of $Q_s=2$ GeV between MATTER and LBT. (Right) Same as the left except with a switching virtuality of $Q_s=\sqrt{2}$ GeV.}
\label{fig:RAA}
\end{figure}

The left panel of Fig. \ref{fig:RAA} explores the description of the $D^0$-meson nuclear modification factor $R_{AA}$ using MATTER down to $Q_s=1$ GeV. The combination of MATTER and LBT, with the switch between the two occuring at $Q_s=2$ GeV, is also presented. Given that the higher twist formalism used in MATTER generates only a few scatterings, these scatterings do not significantly modify the $p_T$-distribution of $D^0$ mesons. This results in a nuclear modification factor being closer to $R_{AA}=1$ than it is to CMS data, as depicted in the left panel of Fig. \ref{fig:RAA}. Using the combination of MATTER and LBT, multiple scatterings occur for partons below $Q_s=2$ GeV according to the LBT formalism. The $D^0$-meson $R_{AA}$ thus obtained is a lot closer to experimental data compared to using MATTER alone. As the mass of the charm quark is smaller than 2 GeV, we have also explored whether changing the virtuality from $Q_s=2$ to $Q_s=\sqrt{2}$ GeV affects our $R_{AA}$ results. These results are presented in the right panel of Fig. \ref{fig:RAA}, where a small effect is seen when reducing $Q_s$ from 2 to $\sqrt{2}$ GeV. Note that the quality of the agreement between the calculated $R_{AA}$ when combining MATTER and LBT and the experimental data partly stems from the improved description of the charm production cross-section in proton-proton collisions when using PYTHIA and MATTER, compared to using PYTHIA alone.   

%%%%%%%%%%%%%%%%%%%%%%%%%%%%%%%%%%%%%%%%%%%%%%%%%%%%%%%%%%%%%%%%
\section{Conclusion}\label{sec:conc}
%%%%%%%%%%%%%%%%%%%%%%%%%%%%%%%%%%%%%%%%%%%%%%%%%%%%%%%%%%%%%%%%

We have presented a first event-by-event calculation of $D^0$-meson $R_{AA}$ using a muti-stage model within the JETSCAPE framework. Very good agreement with CMS data is obtained, partly stemming from the improvement of the p+p baseline when using PYTHIA+MATTER in generating the parton shower in the vacuum. Future work will include determining the $D^0$-meson $v_2$, as well as calculating these observables for bottom hadrons.

%%%%%%%%%%%%%%%%%%%%%%%%%%%%%%%%%%%%%%%%%%%%%%%%%%%%%%%%%%%%%%%%
\section*{Acknowledgments}\label{sec:conc}
%%%%%%%%%%%%%%%%%%%%%%%%%%%%%%%%%%%%%%%%%%%%%%%%%%%%%%%%%%%%%%%%
This research was funded in part by the Natural Sciences and Engineering Research Council (NSERC) of Canada and also by the National Science Foundation (in the framework of the JETSCAPE Collaboration) through award number ACI-1550300. 
 
\bibliographystyle{elsarticle-num}
\bibliography{references} 

\end{document}